\documentclass[10pt]{article}

\usepackage[dvips]{graphicx}
\usepackage{graphics}
\usepackage{enumerate}
\usepackage{amssymb}
\usepackage{amscd}
\usepackage[intlimits]{amsmath}
\usepackage{graphicx}
\usepackage{exscale}
\usepackage{color}
\usepackage{subfigure}

\begin{document}

\title{Remarks on the accuracy of algorithms for motion by mean curvature in bounded domains}

\author{ S.J. Cox$^{\dag\ddag}$, G. Mishuris$^{\dag\ddag}$
\\
$\dag$ Institute of Mathematics and Physics, Aberystwyth University,
UK
\\
$\ddag$ Wales Institute of Mathematical and Computational Sciences}

\date{}

\maketitle

\noindent{\bf Abstract} Simulations of motion by mean curvature in
bounded domains, with applications to bubble motion and grain
growth, rely upon boundary conditions that are only {\em
approximately} compatible with the equation of motion. Three closed
form solutions for the problem exist, governing translation,
rotation and expansion of a single interface \cite{mullins56},
providing the only
benchmarks for algorithm verification. We derive
new identities for the translation solution. Then we estimate the
accuracy of a straightforward algorithm to recover the
analytical solution for different values of the velocity $V$ given
along the boundary. As expected, for large $V$ the error can reach
unacceptable levels especially near the boundary. We discuss factors
influencing the accuracy and propose a simple modification of the
algorithm which improves the computational accuracy.

\section{Introduction}

Foams are ubiquitous in everyday life \cite{WeaireH99}. They are
used daily in the home, in both food and cleaning products.
Moreover, their industrial uses are varied and contribute
significantly to modern
industrial processes. For example, foams are used to separate ores (e.g. zinc,
lead) from the rock in which they are found, and to push oil out of
porous rock. They are also used in the decontamination cleaning of
vessels, and in firefighting. In all these applications, it is the
flow of foam that is driving the process. Therefore, if it were
possible to predict how a foam would behave in this range of
scenarios, when subjected to a complex collection of deformations,
each process could be made more efficient, either in terms of
yield/output or cost-effectiveness.

Aqueous foams and concentrated emulsions, which are very similar to
foams, have peculiar and remarkable properties: they are complex
fluids whose properties lie between the familiar extremes of liquid
and solid. For small strains, a foam behaves as an elastic solid
while at large strains (or strain-rate) a foam moves like a liquid.
They therefore generate a rich range of behaviours, but with a
well-defined local structure that obeys Plateau's geometric laws and
the Laplace-Young law that balances interface curvatures with bubble
pressure differences. Our idea is to use the precise, known,
structure of a foam to predict its rheological response.

One of the leading tools for the analysis of foam structure is the
Surface Evolver developed by Ken Brakke \cite{brakke92}. This
consists of software expressly designed for the modelling of soap
bubbles, foams, and other liquid surfaces shaped by minimizing
energy (such as surface tension and gravity), and subject to various
constraints (such as bubble volumes). Originally designed to model
grain growth in metals, it is easily extended to the case of foams,
and has been used in a number of engineering disciplines to look at,
for example, capillary surfaces, solder joints and fluid behaviour
in microgravity. A surface is represented as a collection of
triangles, so that the complicated topologies found in foams are
routinely handled. In particular, the Evolver can deal with the
topological changes encountered during foam flow.

Here, we concentrate on a two-dimensional
model of a foam, for ease
of analysis. We use the following ``viscous froth model'' (VFM)
\cite{kernwahc03} to examine the evolution of foam films under
shear:
\begin{equation}
  \label{eq:ViscousFroth}
  p_i - p_j   =  \gamma \; \kappa_{ij} + \lambda  v_{ij}
\end{equation}
where $\gamma$ is the surface tension in the films (assumed
constant),
$\kappa_{ij}$ is the curvature of the film separating
bubble $i$ from bubble $j$ and $v_{ij}$ its normal velocity. Each
bubble has a well-defined pressure $p_i$. In essence, the model
augments the (equilibrium) Laplace-Young law with a term
proportional to the velocity; the constant of proportionality
$\lambda$ is a drag coefficient, representing the external
dissipation due to friction with the walls of the container,
allowing the investigation of strain-rate effects. If the external
dissipation is negligible, then we recover the Laplace-Young law for
a foam in equilibrium and a model for quasi-static evolution.

A Surface Evolver simulation with the VFM proceeds in the following
way. Each interface is discretised into short straight segments and
curvatures $\kappa_{ij}$  are calculated pointwise at the
intersections of these segments. Bubble pressures are obtained by
deriving a matrix equation based upon an integral of
(\ref{eq:ViscousFroth}) around each cell. Then the endpoints of the
segments are moved according to the motion equation
(\ref{eq:ViscousFroth}) using a small time step $\Delta t$. Boundary
conditions are applied according to the problem under consideration.

In the ideal model of the evolution of crystalline grains in a
polycrystalline metal, known as normal grain growth, the size of
each grain evolves due to the normal motion of each of its
boundaries \cite{weairem96}. Each boundary has a certain
``mobility'' $\lambda$, and moves in such a way as to reduce the
total perimeter of the pattern, as in foams. However, without area
(volume in 3D) constraints, this is motion by {\em mean curvature}.
As noted above, this is a well-posed limit of the VFM
(see equation (\ref{eq:ViscousFroth})) when bubble pressure differences are
negligible, such as in freely translating films (grain boundaries)
and ordered (hexagonal) foams. The grain growth model requires that
we have $120^\circ$ at vertices, justifying {\em a posteriori} that
assumption in the VFM.

Here, we ask how such a solution can be commensurate with the
boundary of the domain and how well this solution can be calculated
numerically. To the best of our knowledge, answers to this question
are not available in the literature.

\section{Curvature-driven motion of a bounded interface}

\subsection{Problem formulation}

In vector form, the motion of an interface in the model of ideal
grain growth   (\ref{eq:ViscousFroth}) can be described by
\begin{equation}
  \label{Mul_1}
  {\bf v} =  \kappa {\bf n},
\end{equation}
where ${\bf n}$ and ${\bf s}$ are the normal and tangential unit
vectors to the interface (see Fig. 1):
\begin{equation} {\bf
n}=[n_1,n_2],\quad {\bf s}=[n_2,-n_1].  \label{2}
\end{equation}
If the representation of the interface is taken in the form:
\begin{equation} x=x(y,t),  \quad
y\in[\underline{y}(t),\overline{y}(t)],
\label{3}
\end{equation}
then the vector components $n_1,n_2$ are calculated as follows
\begin{equation}
n_1=-\frac{dy}{ds}=-\sin\theta=-\frac{1}{\sqrt{1+(x_y)^2}},\quad
n_2=\frac{dx}{ds}=\cos\theta=\frac{x_y}{\sqrt{1+(x_y)^2}},
 \label{5a}
\end{equation}
where $x_y=dx/dy$, $ds=\sqrt{(dx)^2+(dy)^2}$ and $\theta$ is the
tangential angle to the interface (see Fig. 1).

\vspace{2mm}

\begin{figure}[h]
\begin{center}
\includegraphics[scale=0.7]{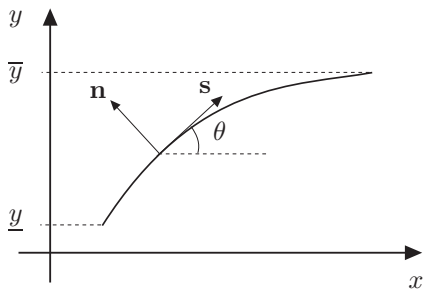}
\end{center}
\begin{picture}(0,0)(-40,-10)
\put(47,-4){$x$} \put(5,24){$y$} \put(5,3){$\underline{y}$}
\put(5,18){$\overline{y}$}\put(13.5,16){${\bf n}$} \put(25,16.5){\bf
s} \put(26.5,11.5){$\theta$}
\end{picture}

\vspace*{-7mm}

\caption{The bounded interface considered here.}
\end{figure}

\noindent Finally, the vector ${\bf v} = [v_1,v_2]$ is the
instantaneous velocity of the point $(x,y)$ lying on the  interface
at time $t$ and $\kappa$ is the curvature of the  interface at that
point:
\begin{equation}
\kappa=\frac{d \theta}{ds}=\frac{-x_{yy}}{\sqrt{(1+(x_y)^2)^3}}.
\label{4}
\end{equation}

Equation (\ref{Mul_1}) can be also written in component form:
\begin{equation} v_n={\bf
v}\cdot{\bf n}=v_1n_1+v_2n_2=\kappa, \label{6}
\end{equation}
\begin{equation}
v_s={\bf v}\cdot{\bf s}=-v_1n_2+v_2n_1=0. \label{7}
\end{equation}
In this paper we will consider only Mullins' solution
\cite{mullins56}, also known as the grim reaper, (see below) which
is symmetrical with respect to the $x$-axis. Taking into account the
direction of the interface motion, we can assume in what follows
that:
\begin{equation} n_1<0,
\hspace{2mm} n_2>0,\hspace{2mm} x_y>0, \hspace{2mm} 0<\theta<\pi/2,
\hspace{2mm} x_{yy}>0,\hspace{2mm} \kappa<0,
 \label{signs_1}
\end{equation}
\begin{equation}
v_n<0, \hspace{2mm}  (v_2<0, \hspace{2mm} v_1>0). \label{signs_2}
\end{equation}
Equation (\ref{6}) is widely discussed in the literature
\cite{mullins56,PMV}, while the second equation (\ref{7})  is
somehow usually forgotten in this context. If one is only interested
in reconstruction of the interface position at any time step an
approach based only on equation (\ref{6}) is sufficient. However, if
it is required to control the position of each tessellation point along
the interface, as in the case of numerical computation, then both
equations are equally important.
Note that there has previously been an
attempt to control  both the velocity components in a specific way in
\cite{Grassia1}. 
Equation
  (\ref{7})
allows us to find relation between the two unknown components of the
velocity vector ${\bf v}$ and the normal vector ${\bf n}$ in the
form: 
\begin{equation}
n_2=\frac{v_2}{v_1} n_1. \label{h1}
\end{equation}
This allows us to eliminate components of the normal vector ${\bf
n}$ from equation (\ref{6}) to give:
\begin{equation}
-\left( v_1+\frac{v_2^2}{v_1}\right)=\frac{d\theta}{dy}. \label{9a}
\end{equation}

\subsection{Mullins' solution for translation revisited}

Let us assume that the interface conserves its shape but moves in the
$x$-direction with a constant speed $V$. We consider two points $A$
and $C$ having the same $y$-coordinate $y=y_0$ at two consecutive time steps
$t_0$ and $t_0+dt$ (see Fig. 2). It is clear that these two
points correspond to two different material points. Namely, there
exists a point $B$ on the interface at time $t_0$  which moves according to
the curvature law (\ref{Mul_1}) to the point $C$ for an infinitesimally
small time step $dt$. If the coordinates of the point $A$ are
$(x_0,y_0)$ then the coordinates of  $B$ and $C$ can be
written  $(x_0+dx,y_0+dy)$ and $(x_0+V dt,y_0)$.

\begin{figure}[h]
\begin{center}
\includegraphics[scale=0.6]{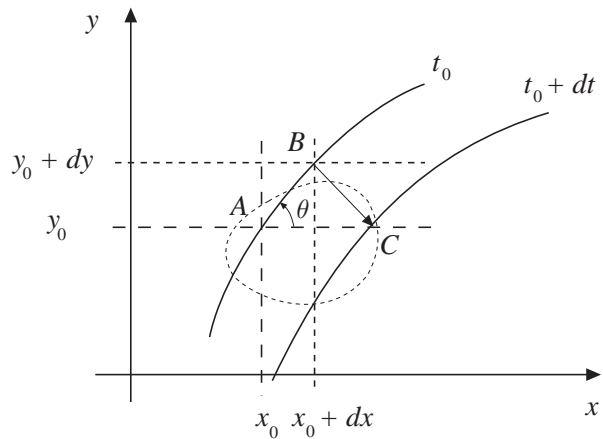}\hspace{5mm}
\end{center}
\begin{picture}(0,0)(-40,-10)
 \put(26.2,19.5){$\theta$}
\end{picture}

\vspace{-10mm} \label{Fig.2}
\caption{Interface under translational
motion at two consecutive instants in time $t_0$ and $t_0+dt$.}
\end{figure}

Note that
\[ 
\overline{BC}={\bf v}(x_0+dx,y_0+dy)dt=
{\bf v}(x_0,y_0)dt+O(dt ds).
\]
On the other hand, $|BC|=V\sin\theta dt$, $\tan\theta=v_1/|v_2|$. As
a result one can conclude:
\begin{equation}
V=\frac{1}{v_1(y)}\left(v_1^2(y)+v_2^2(y)\right), \label{Mul_10}
\end{equation}
in some interval $y\in[0,h]$. This relation
immediately follows in the case of the translation movement of the
interface in $x$-direction from (\ref{9a}).

Now, to reconstruct the solution obtained by Mullins \cite{mullins56} it
is sufficient to substitute (\ref{Mul_1}) into (\ref{9a}) to have:
\begin{equation}
\pi/2-Vy=\theta. \label{Mul_2}
\end{equation}
Here we have taken into account the second of the two symmetry conditions at the point
$y=0$:
\begin{equation}
v_2(0)=0,\quad \theta(0)=\pi/2. \label{Mul_3}
\end{equation}
Equation (\ref{Mul_2}) can be written in the form
\begin{equation}
\cot Vy=y_x, \label{Mul_4}
\end{equation}
which after direct integration leads to Mullins' solution
\cite{mullins56}:
\begin{equation}
x(y)=x(0)+Vt-\frac{1}{V}\log\cos(Vy),\quad
y\in[0,h]. \label{Mul_5_0}
\end{equation}
where $x(0)$ is the arbitrary initial
position of the centre of the interface. This solution exists only
under the condition $h<h_{max}$,
where
\begin{equation}
h_{max}=\frac{\pi}{2V}. \label{Mul_5a}
\end{equation}
Note also that the angle $\theta$ defined by such a solution
monotonically decreases in the interval $y\in(0,h)$ ($h<h_{max}$),
taking values
\begin{equation}
\theta(y)\in(\theta_{min},\pi/2),\quad
\theta_{min}=\theta_{min}(h)=\pi/2-Vh.\label{Mul_5aa}
\end{equation}
It is now possible to write analytical representations of all
problem variables in the interval $y\in(0,h)$:
\begin{equation}
\label{Mullin_variables} v_1=V\cos^2 Vy,\quad
v_2=-\frac{1}{2}V\sin2Vy,\quad  n_1=\sin Vy,\quad n_2=\cos Vy,\quad
\kappa=V\cos Vy.
\end{equation}
Note that the first symmetry condition (\ref{Mul_3}) has not been used but the
reconstructed Mullins solution (\ref{Mul_5_0}) satisfies it
automatically. It exhibits the following
asymptotics near the symmetry axis:
\begin{equation}
\label{Mullin_est2}
x(y)=x(0)+Vt+\frac{1}{2}
Vy^2+O(y^4), \quad y\to 0;
\end{equation}
thus near $y=0$ the interface is close to a
parabola. Near the other end of the reaper (in the case of the maximal
thickness $h=h_{max}$), the following asymptotic estimate can be
obtained:
\begin{equation}
\label{Mullin_est2b} x(y)= -\frac{1}{V}\log(h_{max}-y)+O(1), \quad
y\to h_{max},\quad \mbox{or}\quad
y-h_{max}\sim -de^{-Vx}, \quad x\to +\infty,
\end{equation}
where $d>0$ is a constant.

\noindent {\bf Remark}: Note that relation (\ref{Mul_10}) also
represents the boundary condition for any moving interface
whose upper point lies on the line $y=\overline{y}=h$ and which moves
in the $x$-direction with velocity $V$. Moreover, the velocity can be in
that case a function of time $V=V(t)$:
\begin{equation}
V(t)v_1(h,t)=v_1^2(h,t)+v_2^2(h,t). \label{Mul_11}
\end{equation}
Substituting (\ref{Mul_1}) into (\ref{Mul_11}) such a boundary
condition can be equivalently rewritten in other forms:
\begin{equation}
\kappa(h,t)=V(t)n_1(h,t), \quad \mbox{or}\quad
x_{yy}(h,t)=V(t)\left( 1+(x_y(h,t))^2\right ). \label{BC}
\end{equation}
Note that we have not used (\ref{Mul_5_0}) to define
(\ref{BC}).

\subsection{Arbitrary instantaneous solution of equations (\ref{6}) -
(\ref{7}) in bounded domain.} \label{sect_2.3}

Let us consider any instantaneous solution of the equations
(\ref{6}) - (\ref{7}) with the prescribed boundary condition
(\ref{Mul_11}). Effectively this means that, for a particular time
$t$, the end points of the interface
$y=\underline{y}(t)$ and $y=h$ are defined and the velocity
components $v_1(y,t)$ and $v_2(y,t)$ are known functions of the
variable $y$, while the problem is now to determine, using this
information, the position of the interface in space variables
$(y,x)$.

Let us introduce a function which in what follows is considered known:
\begin{equation}
F(y)= v_1(y)+\frac{v_2^2(y)}{v_1(y)}>0,\quad
y\in(\underline{y},h).\label{10}
\end{equation}
Note that the condition (\ref{Mul_10}) may be not
valid at all inside the interval $y\in(\underline{y},h)$ as it was for  Mullins' solution;
as a result, $F(y)$ is not a constant, in general.
Equation (\ref{9a}) can be integrated to give:
\begin{equation}
-\int\limits_{\underline{y}}^y F(\xi)d\xi=\theta(y)-\underline{\theta}.
\label{10a}
\end{equation}
Here, recall that $F$ depends upon time $t$, so that the constant of integration
$\underline{\theta}=\underline{\theta}(t)$ and 
$\underline{y}=\underline{y}(t)$. The equation (\ref{10a}) should
be considered together with equation (\ref{h1}) which, in this case,
takes the form:
\begin{equation}
\frac{dx}{dy}=-\frac{v_2}{v_1}=\cot \theta,\label{h2}
\end{equation}
or
\begin{equation}
x(y)=\underline{x}-\int\limits_{\underline{y}}^y\frac{v_2(\xi)}{v_1(\xi)}d\xi.\label{h2a}
\end{equation}
 Equations (\ref{10a}) and (\ref{h2}) together indicate that the
functions $v_1$ and $v_2$ cannot be chosen arbitrarily to satisfy the
vectorial equation (\ref{Mul_1}) as one might expect.
Instead, the following identity has to be satisfied:
\begin{equation}
\arctan\frac{v_1}{v_2}=\int\limits_{\underline{y}}^yF(\xi)d\xi+\underline{\theta},\label{h3}
\end{equation}
or writing $w=v_1/v_2$, equation (\ref{h3}) becomes:
\begin{equation}
\frac{w^\prime}{1+w^2}=\frac{(1+w^2)v_2}{w},\label{h4}
\end{equation}
which leads to the following identity
valid within the entire interval $y\in(\underline{y},h)$:
\begin{equation}
v_1^2(y)=-v_2^2(y)\left(\frac{1}{2\int
\limits_{\underline{y}}^yv_2(\xi)d\xi+c}+1\right),\label{h7}
\end{equation}
where the constant of integration clearly depends on time too, $c=c(t)$. Note that any solution of
equations (\ref{6}) - (\ref{7}) satisfies this additional relation,
which makes sense only under the additional constraint:
\begin{equation}
-1\le
2\int\limits_{\underline{y}}^yv_2(\xi)d\xi+c \le
0.\label{restriction_1}
\end{equation}
As $0<\theta<\underline\theta $, one can also deduce another
condition which has to be true for any admissible velocities:
\begin{equation}
0\le \int\limits_{\underline{y}}^yF(\xi)d\xi\le \underline{\theta}.
\label{restriction_2}
\end{equation}

Note that the constant $\underline{x}$ in (\ref{h2a}) is arbitrary
(it changes only the position of the interface in the $x$-direction
and does not influence any other variables). To determine the other
constants$\underline{\theta}(t)$
and $c(t)$, we need to use the boundary conditions at the ends of the
interface. Thus, condition (\ref{Mul_11}) together with (\ref{h7})
leads to
\begin{equation}
v_1(h)=V\left(1+c+2I_2\right), \quad
v_2(h)=-V\sqrt{-\left(1+c+2I_2\right)\left(c+2I_2\right)}.
\label{super_condition}
\end{equation}
where we denote
\[
I_2\equiv I_2(t)=\int\limits_{\underline{y}(t)}^hv_2(\xi)d\xi.
\]
If the boundary condition on the other end is given in the form
\begin{equation}
\theta(\underline{y}(t))=\underline{\theta}(t),
\label{other condition}
\end{equation}
then all the constants have been defined. Such an instantaneous
solution, assuming that the functions $v_1(y)$ and $v_2(y)$ satisfy
equation (\ref{h3}),
 conditions (\ref{super_condition}) and restrictions (\ref{restriction_1}) and (\ref{restriction_2}),
can be realised during the interface evolution at some step.

In the case of the symmetrical solution, where both symmetry
conditions (\ref{Mul_3}) and the additional condition $v_1(0)=W>0$ have to be satisfied, one can show that
\begin{equation}
c(t) \equiv 0,\quad v_2(y)\sim
-W^2y,\quad y\to0.\label{h9}
\end{equation}
Note here that the value $W=W(V,h)$ should be found from the
constructed solution and is not an additional (arbitrary or given)
constant.

Finally, both restrictions (\ref{restriction_1}) and
(\ref{restriction_2}) should be valid for the symmetrical interface:
\begin{equation}
\int\limits_{0}^hv_2(\xi)d\xi\ge-1/2,\quad \int\limits_{0}^h
F(\xi)d\xi\le \pi/2.\label{h12}
\end{equation}
Note that the tangential angle $\theta$ for this solution
is a monotonically decreasing function in the interval $y\in(0,h)$
so that
\begin{equation}
\theta(y)\in(\theta_{min},\pi/2),\quad
\theta_{min}=\pi/2-\int\limits_{0}^h F(\xi)d\xi. \label{Mul_5b}
\end{equation}
It is straightforward to see that Mullins' solution
(\ref{Mullin_variables}) satisfies all these relationships with
$c(t)=0$,
$\underline{\theta}(t)=\pi/2$ and $W=V$, as expected. In the
Appendix, we construct analytical examples of symmetrical
instantaneous solutions which are different from Mullins'.
Those solutions are not, generally speaking, steady-state ones. This
means that they can be reached at some time step $t$, given the
interface boundary velocity $V(t)$ and the position of the ends
$h(t)$), but all these parameters may later change with time. What
is extremely interesting about these solutions is that some of them
are well-defined for any velocity $V>0$ and an arbitrary position of
the boundary $y=h$. This shows a ``rich behaviour" of possible
instantaneous solutions. It is also clear that there is an infinite
number of admissible instantaneous solutions. Some of them can be
realised during some specific non steady-state interface motion. For
example,  {\it any} instantaneous solution obtained during a
numerical computation, for any particular time step, boundary
velocity and topology, has to satisfy all the relations (\ref{10}) -
(\ref{other condition}). This will allow us to use the relations as
indicators of the accuracy of computations. Moreover, they could
provide a means to improve the accuracy of the algorithms.

Note also that a family of arbitrary (asymmetric) instantaneous
solutions constitutes an even larger set in comparison to the
symmetric case. In fact, the
 family of symmetrical solutions has one degree of freedom (since the constant $c(t)$
 is equal to zero in this case) and correspondingly one less boundary
condition (compare (\ref{other condition}) and (\ref{Mul_3})).
Moreover, in the context of further applications of this result to a
given algorithm, where the angle-type boundary condition has to be
preserved at the interface intersection point, it is worth
mentioning that the boundary condition (\ref{other condition}) is
therefore more important for application than the symmetry
condition. On the other hand, symmetrical instantaneous solutions
can also be considered a subset of the asymmetric solutions if one
considers the interval $(h_0,h)$ instead of $(0,h)$; ($0<h_0<h$).
This idea has been exploited previously
 in \cite{Grassia2}.

\section{Numerical Simulations}

\begin{figure}
\centerline{
\includegraphics[scale=0.65]{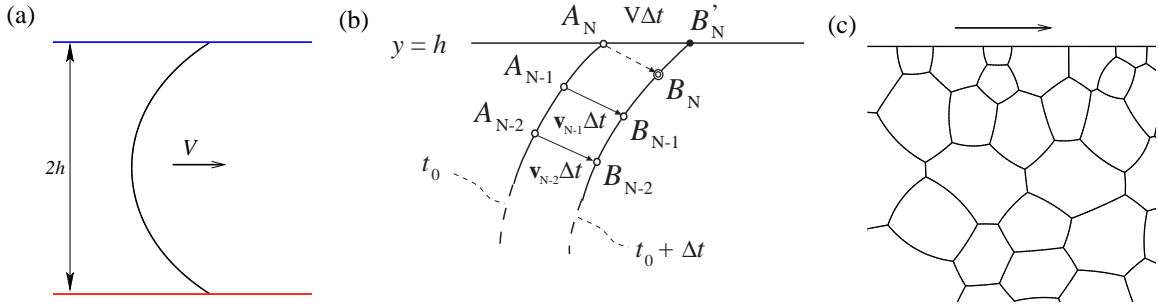}
} \caption{\small (a) The test problem considered here consists of a
single interface that is sheared symmetrically by translating the
boundaries. The shape should correspond to Mullins' solution. (b)
Standard algorithm for implementation of the boundary condition at
each time step $\Delta t$. $A_j$ and $B_j$ are the respective
tessellation points on the interface at times $t_0$ and $t_0+\Delta t$.
(c) Example of a multi-bubble foam simulation, for which the
numerical procedure developed here will be of use.}
\label{fig:sheared211}
\end{figure}

To indicate the computational inaccuracy, we discuss
Mullins' solution for a symmetrical reaper, for which all quantities
are known in closed form (see (\ref{Mul_2}) and remarks thereafter),
and
compare it with the result of a numerical computation using a
simple algorithm (implemented in the Surface Evolver). This takes
the form of a single interface separating two bubbles of equal pressure
being pulled at a velocity $V$ at each boundary (fig.
\ref{fig:sheared211}).

The numerical procedure can be briefly described as follows. We
start from a straight (vertical) line joining the two walls a
distance $2h=2$ apart. This is subdivided into $2^5$ short elements
(``edges'' which meet at points). A time step $\Delta t =
1\times10^{-5}$ is chosen for the computations and bounds on the
possible length $L$ of each edge ($0.01 \le L \le 0.05$). The
algorithm proceeds as follows: (i) each boundary point is moved in
the $x$ direction a distance $V \Delta t$; (ii) the curvature at
each point that is not on the boundary: $ \kappa ={\bf F} \cdot {\bf
n}/{\bar{L}} $ where $\bar{L}$ is the average length of the
neighbouring edges and ${\bf F}$ is the negative energy (perimeter)
gradient \cite{brakke92}; (iii) each point
that is not on the boundary is then moved according to $ \Delta {\bf
r} = \Delta t \; \kappa \; {\bf n}$ (fig. \ref{fig:sheared211} b)).
This procedure (one ``step'') is repeated until  the centre-point of the interface moves less than a critical value ($1\times
10^{-8}$). Every 20 steps we check the edge length bounds and add or
remove edges as necessary. Note that
this standard algorithm preserves a reasonable restriction on
the length of the edges; however, it works in a way that
does not guarantee equal length edges.

Note that this choice of the
parameters for numerical simulation is standard and allows us to
obtain acceptable accuracy in reasonable computational time
\cite{cox05}. On the other hand, when one computes the dynamics of
foams with many bubbles, the total computational error accumulates.
Therefore information about the error is crucial, since it gives us
a lower bound for the total computational error.

Two important observations illustrating the weakness of the
algorithm should be noted here:
\begin{itemize}
\item
The density of the tessellation points near the symmetry axis
($y=0$) increases with each time step (fig. \ref{fig:sheared211}
b). Since the time step is constant, this may lead to failure of the stability condition for
the linearized finite difference (FD) scheme applied to the nonlinear
parabolic equation (\ref{Mul_1}) due to this algorithm.
\item
The opposite effect occurs near the external boundary $y=h$.
However, the situation here is even much worse. In fact, there is
not enough information to reconstruct the curvature and the unit
vector at a point $A_N$ lying on the boundary and the algorithm, in
fact, simply eliminates it. It creates instead the point
$B_N^\prime$ along the boundary which should be the next point
$B_{N+1}$ (fig. \ref{fig:sheared211} b)). See \cite{Grassia1} for more details.
\end{itemize}

First we should underline here that the computations were stable
(the stable steady-state regime has been reached) for every value of
the external velocity $V$ under consideration. Note that in our
computations at high $V$, the number of tessellation points has
increased from $2^5$  to about 220 at steady-state. As
expected, the worst situation (in the sense of computational time) occurs
for the largest value of the velocity, $V=1.560796$, which is
slightly less that the critical value $V_{cr}=\pi/(2h)$ predicted by
the analytical solution \cite{mullins56}. In numbers, it takes about
1 hour to complete the computation in the case $V \le 0.5$, rising
to over 48 hours for $V=1.560796$ (SGI Altix Itanium 2, 1.5GHz). The
algorithm could not reach the steady-state regime at all for
$V>1.560796$, and lost physical meaning since the interface had a branch
lying outside the external boundary $y=h$. All this illustrates that
the existing algorithm is well organised and works according to
expectations but it is naturally sensitive to the value of the
boundary velocity $V$. Thus it makes sense to ask about algorithm
accuracy for a specific velocity versus space and time steps.

\begin{figure}[h]
\begin{center}
\subfigure[]{
\includegraphics[scale=0.85]{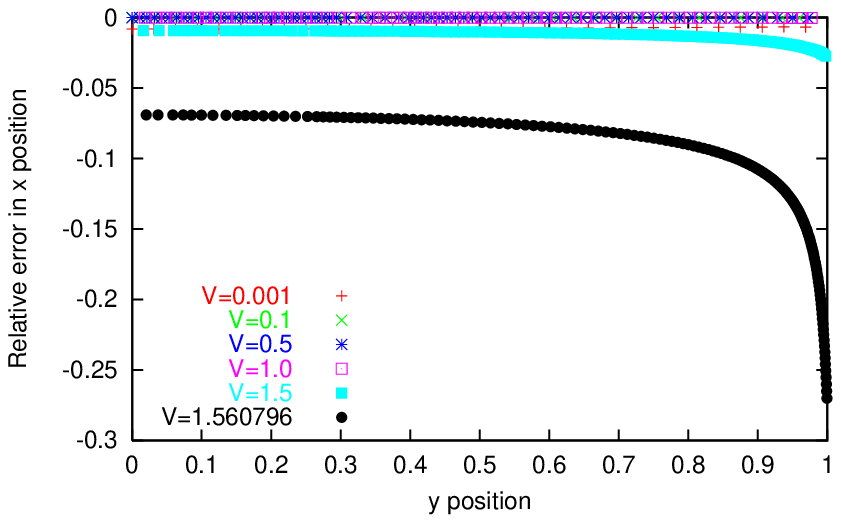}
}
\subfigure[]{
\includegraphics[scale=0.85]{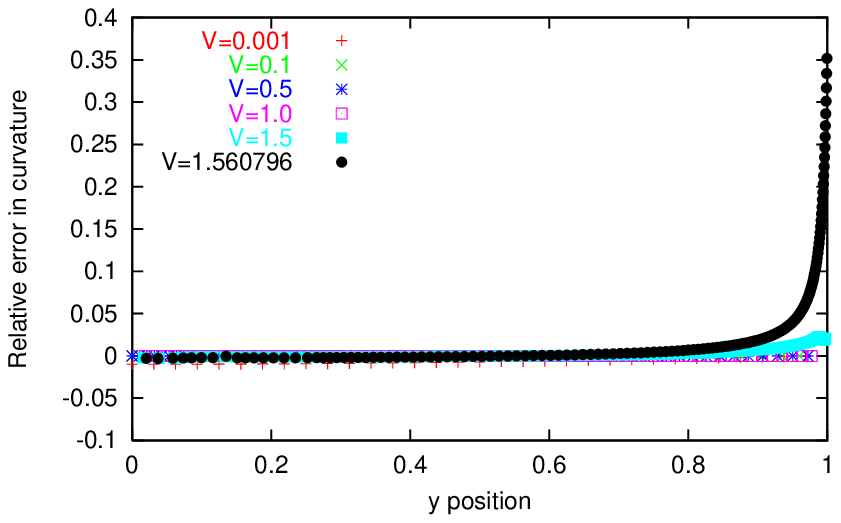}
}
\subfigure[]{
\includegraphics[scale=0.85]{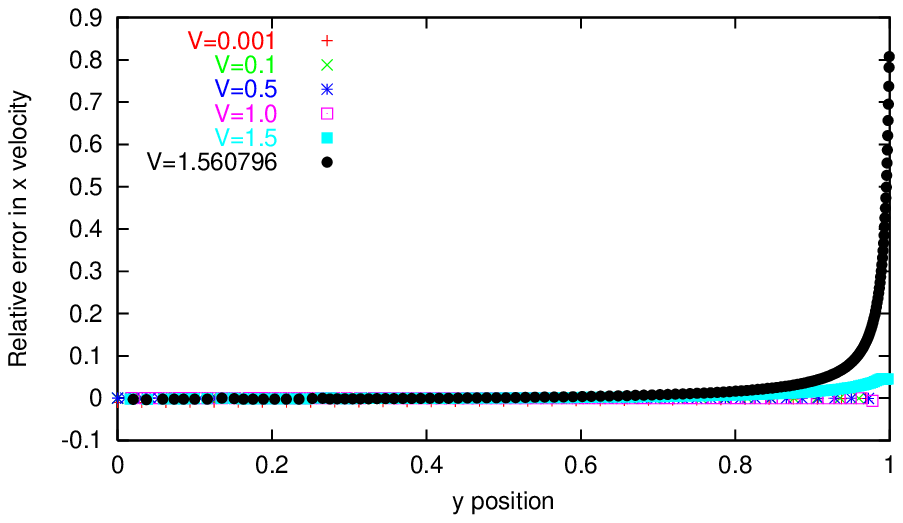}
}
\subfigure[]{
\includegraphics[scale=0.85]{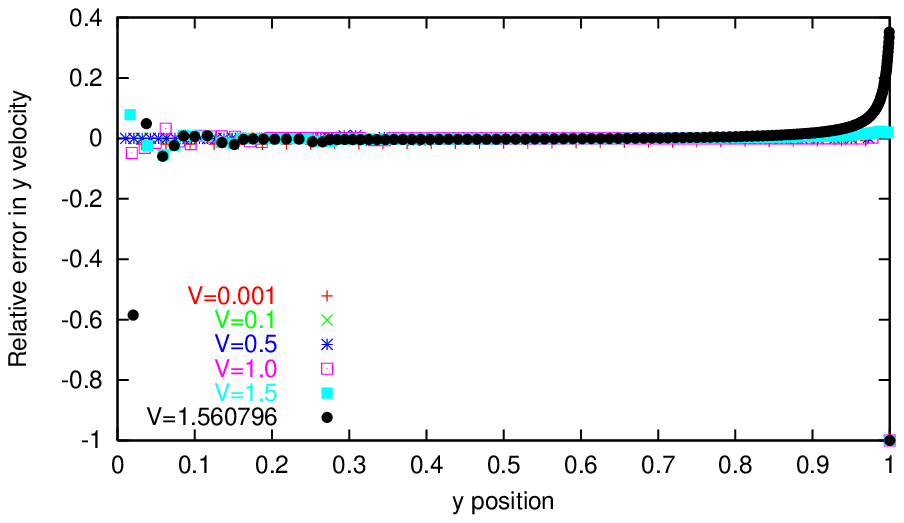}
} \caption{Relative error, compared to Mullins' solution (\ref{Mul_5_0})
and (\ref{Mullin_variables}), in (a) the position of the interface
$x(y)$, (b) its curvature $\kappa(y)$, and  the velocity components
(c) $v_1(y)$ and (d) $v_2(y)$, for different velocities $V$ of the
external boundary.}
 \label{fig:firstresults1}
\end{center}
\end{figure}

As the exact analytical solution to the Mullins problem is known, we can
estimate errors in the computations for all the
physical and geometrical quantities: position of the interface $x(y)$,
curvature $\kappa(y)$ and the velocity components $v_1(y), v_2(y)$.
Corresponding relative errors for all solution parameters are presented in fig.
\ref{fig:firstresults1} for different applied velocities.

As expected, the least accurate solutions are those for the largest
velocity $V=1.560796$. The error can be as large as $90\%$ near the
boundary ($y=h$) in the $x$ component of velocity $v_1$ and $60\%$
near the symmetry axis for the $y$ component.
The latter error is naturally
related to the fact that the value of the velocity is equal to zero
at the symmetry point. However, the drastic difference in value for
this {\it numerical noise}, and its distance from the axis, indicates
that at the larger values of $V$ it exceeds reasonable
expectations and is really related to the computational accuracy. 

One can also consider that the error
near the external boundary is due to parametrisation of the numerical
solution in $y$ rather than $x$, but the standard
numerical algorithm tries to preserve edge lengths. Moreover, the algorithm introduces new points in a regular fashion.

Thus both the errors (near the interface
ends) are a consequence of the phenomena discussed in the
 two observations above. As $V$ decreases, the
accuracy increases for given bounds on the edge lengths $L$.

At first glance, it would appear that the position of the interface, $x(y)$, should be
computed with better accuracy than all other solution parameters, which
are, in fact, results of some derivative procedure. However, our
computations show that this is not in the case and the relative
error for $x(y)$ varies from $6\%$ to $28\%$ for the velocity
$V=1.560796$ while the curvature error is lower. Note also that the
error for smaller velocities reaches a few percent
near the boundary or symmetry axis.

The maximal absolute errors for all solution parameters  mostly
appear near the external boundary $y=h(=1)$. This highlights that
the implementation of the boundary condition in the existing
algorithm cannot be considered as sufficient and should be improved.

\begin{figure}[h]
\begin{center}
\subfigure[]{
\includegraphics[scale=0.85]{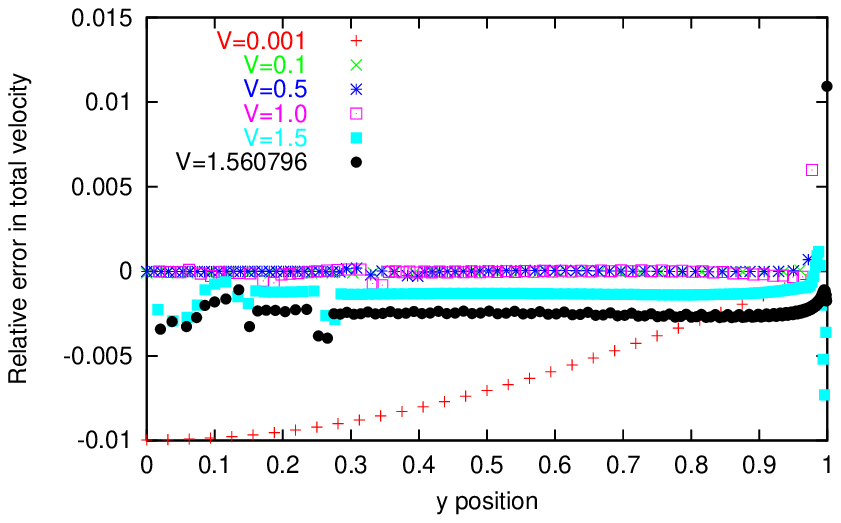}
}
\subfigure[]{
\includegraphics[scale=0.85]{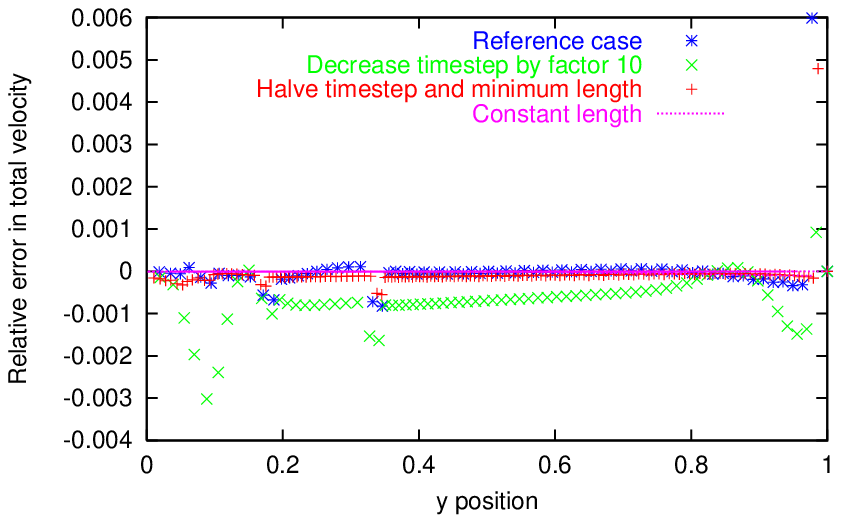}
} \caption{Relative error for the characteristic relation
(\ref{Mul_10}) for (a) varying velocity $V$, and (b) varying
numerical parameters ($\Delta t, L$). The best accuracy is obtained
by keeping the line segments of equal length. }
 \label{fig:firstresults_2}
\end{center}
\end{figure}

Moreover, in the case of Mullins' solution an additional simple local
indicator defined by identity (\ref{Mul_10}) (independent of the
integration of the solution variables) could equally be considered. It is clear from the results presented in fig.
\ref{fig:firstresults_2}(a), that the error in this condition is not
localized near the ends of the interface, as one might expect from the
above.
 Moreover, this ``internal error" is present for all
values of $V$ and is comparable with that near the interface ends.

To investigate accurately this ``internal'' error we repeated the
computations for a specific velocity $V=1$ and decreased both the time increment, $\Delta t$, and the minimum edge length, $L_{\rm min}$ (fig.
\ref{fig:firstresults_2}(b)).
 This has improved the quality of the computations, but
there is still an error at some internal points of the interval that
is comparable with the error near the ends. Unfortunately, it
considerably increases the computational time by several hours. The
obvious route of decreasing $\Delta t$ but fixing $L_{\rm min}$, to
ameliorate this effect, leads to greater error in the solution (fig.
\ref{fig:firstresults_2}(b)). The accuracy of the solution can be
improved internally by making the line segments of equal length (fig.
\ref{fig:firstresults_2}(b)) \cite{Grassia1}, although this doesn't affect the error at the boundary.

Possible sources of the ``internal" error include: a) non-optimal
distribution of the tessellation points along the interface after some
time; b) imperfections in the correction procedure (which adds and
eliminates points from the interface at some prescribed time); c)
point-to-point error variation related to the fact that the
``diffusion"-type coefficient changes from point to point along the
interface (recall that equation (\ref{Mul_1}) is a nonlinear parabolic
equation which is solved by a direct FD scheme with a fixed time step).

\begin{figure}[h]
\begin{center}
\subfigure[]{
\includegraphics[scale=0.85]{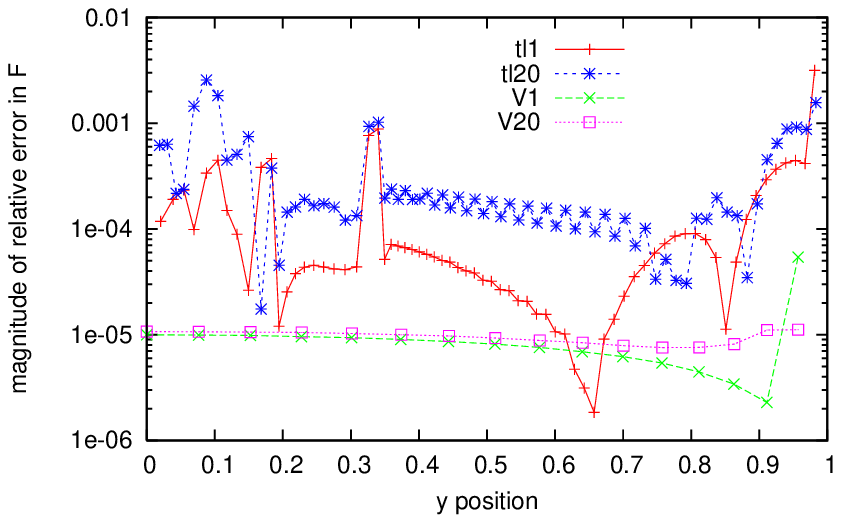}
}
\subfigure[]{
\includegraphics[scale=0.85]{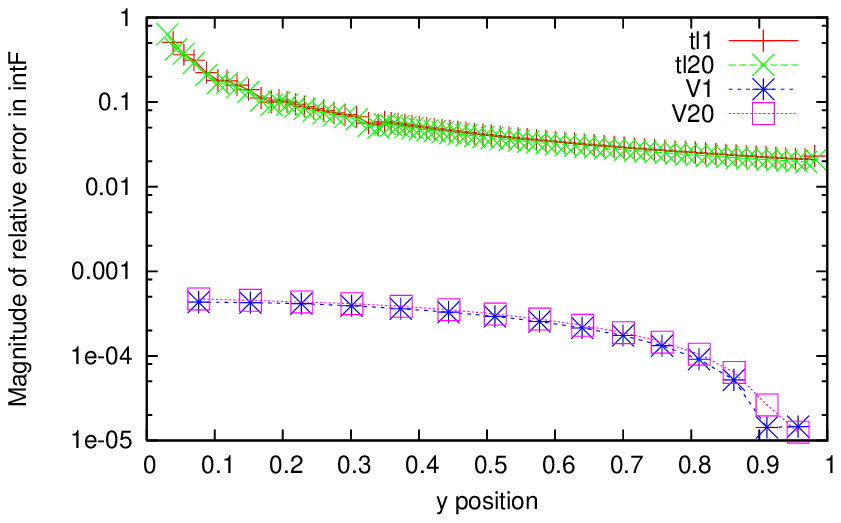}
}
\subfigure[]{
\includegraphics[scale=0.85]{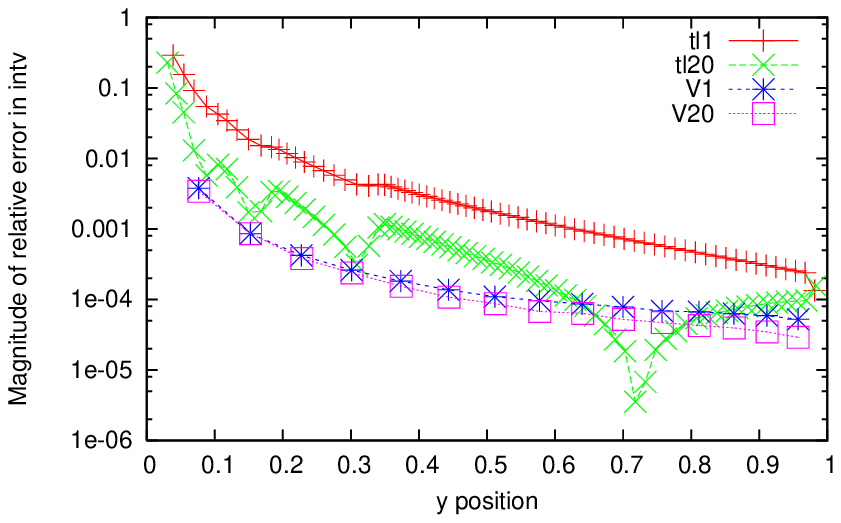}
}
\subfigure[]{
\includegraphics[scale=0.85]{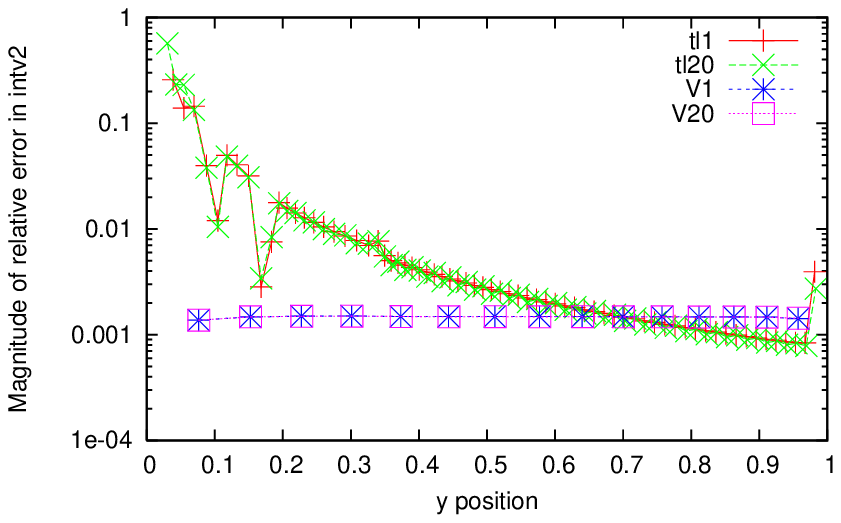}
} \caption{Relative errors in the computations shown by the integral
measures for unit velocity of the external boundary, $V=1.0$,
obtained with four different computational strategies: the standard
redistribution of the tessellation points, tl1 and tl20; uniform
length segment redistribution, V1 and V20, at every time step and
every $20^{\rm th}$ step, respectively. (a) the function $F$ defined
in (\ref{10}) (as in Fig. \ref{fig:firstresults_2}); (b,c,d) the
three general internal measures; for the notation used in the graphs
see Appendix \ref{app:notate}. All integrals  were been computed
with the trapezium rule.}
 \label{fig:firstresults_3}
\end{center}
\end{figure}

In the last computation in figure \ref{fig:firstresults_2}(b), for
the line segments of equal length, we have redistributed points to
make the segments equal at every time step, but note that this length may change in time. Apart from the fact that the
number of tessellation points is smaller than for the standard
algorithm, such a comparison is not absolutely fair as the
redistribution in the standard algorithm was done every 20 time
steps. To discover if there is an effect of redistribution frequency
on the accuracy, we have tested these two algorithms under the same
strategy by redistributing the points (in a different way) every
time step and every $20^{\rm th}$ time step. The error in the
function $F$ defined in (\ref{10}), which is a constant in the case
of Mullins solution, are presented in \ref{fig:firstresults_3}(a).
It is evident that the standard algorithm is quite sensitive to the
chosen strategy. For a given position of the points on the
interface, the relative error may differ by as much as two orders of
magnitude, whereas this is not the case for the equal segment
strategy. In this case, only near the external boundary is there
some small fluctuation in the accuracy. Comparing the two
redistribution algorithms for the same frequency of redistribution,
the largest error always appears in the case of the standard
algorithm -- by up to two orders of magnitude -- despite the fact
that the number of tessellation points was greater. Moreover, in the
standard algorithm this error is irregularly distributed along the
interface. Recall that the number of tessellation points in the
standard algorithm changes during the computations from $2^5$
initially to around $150$ (for $V=1$) in the steady-state regime,
while the number of the points in the second (equal length)
algorithm remains constant. Therefore the computational time for the
second algorithm was less by a factor of approximately two. On the
other hand, the difference in the computational time between the
different frequencies for redistribution  for the equal length
algorithm was only a few percent. This indicates the further
possibility to optimise this algorithm by redistributing points
every $M$ time steps. It is clear that $M=M(V)$ and this needs
further investigation \cite{Grassia1}. 

Note that the function $F$ from
(\ref{10}) can be used as an indicator of the accuracy of the
computation only for the Mullins solution. However, there are three
universal indicators which can be helpful to estimate the accuracy
for any computations. Namely,  the relative errors of the numerical representations
of the identities (\ref{10a}), (\ref{h2a}) and (\ref{h7}). The
respective results are shown in \ref{fig:firstresults_3}(b)-(d)).

As in the case of the
specific indicator discussed above (Fig. \ref{fig:firstresults_3}(a)),
all general indicators presented  in Fig. \ref{fig:firstresults_3}
(b) -- (d) indicate that the equidistant distribution of the
tessellation points is much better than the standard algorithm,
regardless of the chosen strategy. Moreover, even near the symmetry
point, $y=0$, where the value of the indicators all tend to zero and have a large influence on the relative errors, the accuracy of
the computations for the second algorithm is extremely
high. This is not the case for the standard algorithm.

\begin{figure}[h]
\begin{center}
\subfigure[]{
\includegraphics[scale=0.85]{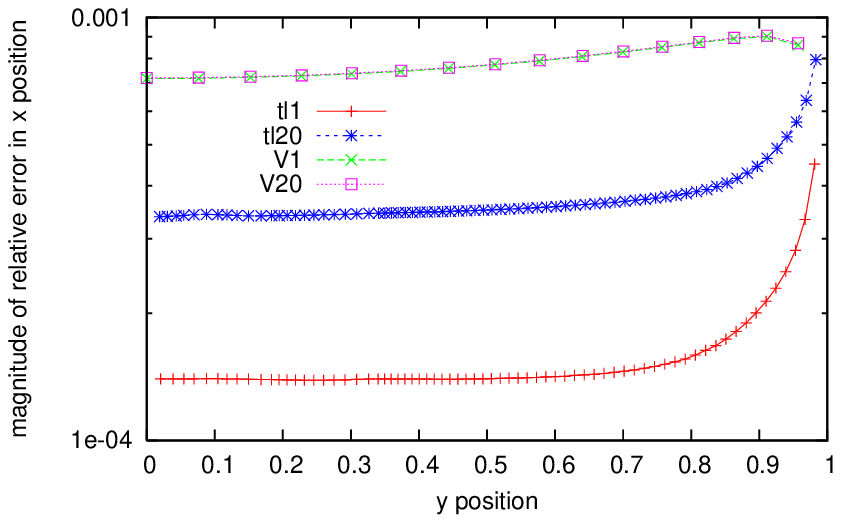}
} \subfigure[]{
\includegraphics[scale=0.85]{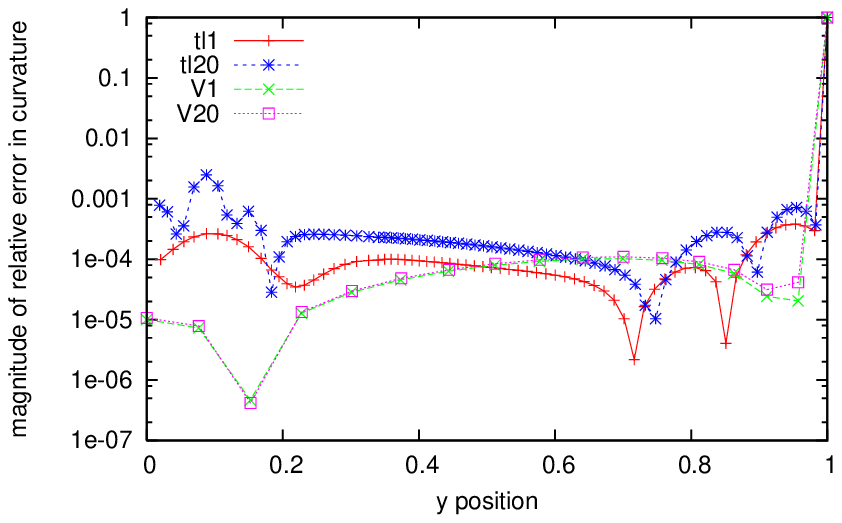}
}
\subfigure[]{
\includegraphics[scale=0.85]{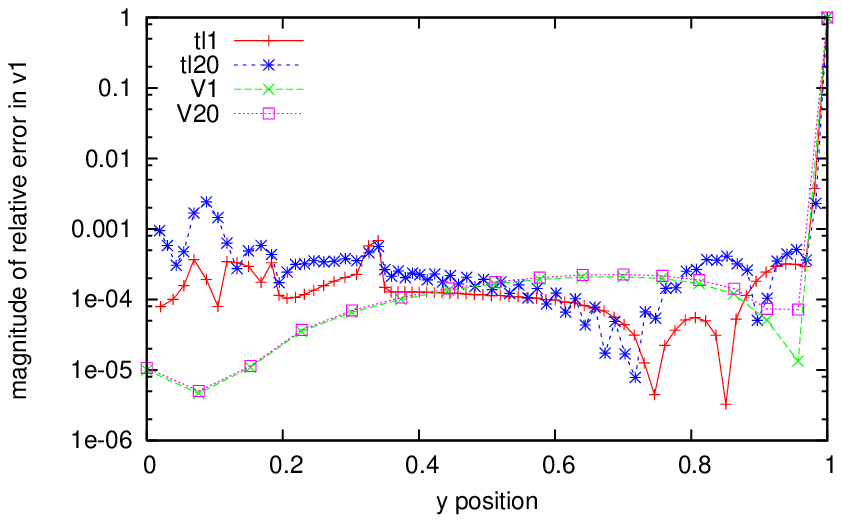}
} \subfigure[]{
\includegraphics[scale=0.85]{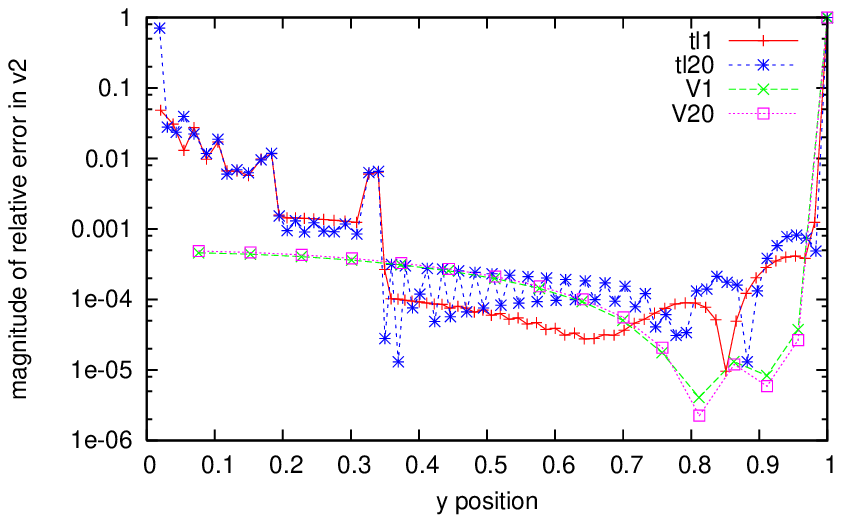}
}
\caption{Relative error in (a) the
position of the interface $x(y)$, (b) the curvature $\kappa(y)$, and  (c) and (d)
the velocity components  $v_1(y)$ and $v_2(y)$, for velocity  of the
external boundary $V=1$ and different numerical algorithms. }
 \label{fig:firstresults_4}
\end{center}
\end{figure}

In Fig. \ref{fig:firstresults_4}, the relative errors of the
solution variables are presented for both algorithms: the standard
one and the equidistant distribution. The result shown in Fig.
\ref{fig:firstresults_4}(a) looks surprisingly at first glance:
although the accuracy of the computations performed with these two
algorithms is of the same order and the error related to the new
algorithm is distributed more uniformly, it appears that the
accuracy of the standard algorithm is better than the
equal-segment-length algorithm, at least with respect to the
accuracy of the position of the interface. However, this not in the
case. In fact, as was shown above, the computational error for the
standard algorithm is redistributed along the interface irregularly
whereas that for the equal-segment algorithm is practically uniform.
As a result, the criterion to stop the iteration process to find the
steady-state solution works differently for the two algorithms. The
prescribed maximal growth $10^{-8}$ in each time-step, measured on the axis of symmetry, is
reached more quickly for the new algorithm. This is the second
reason (together with number of tessellation points) why this
algorithm is faster. If one were to run both algorithms for the same
time, or for the same number of iterations, and compare the
corresponding results, the discussed paradox should not appear and
the new algorithm always provides better accuracy by as much as two
orders of magnitude.

However, for the accuracy of other problem variables, the interface
curvature, $\kappa$, and the interface velocities, $v_1$ and $v_2$,
cf. in Fig. \ref{fig:firstresults_4} (b)-(d), the new algorithm is
more accurate, notwithstanding the above argument.

Finally, let us stress again that the proposed three indicators are
more versatile measures than a
comparison of numerical steady-state solution with the analytical
one, since the latter comparison includes an additional error
related to the determination of the steady-state regime, while the
indicators show us accuracy of the solution at any time step.

\section{Discussion and Conclusions}

All these results clearly indicate that the existing algorithm
should be used with caution, especially when investigating foam
behaviour near the critical velocity. Moreover, when there are many
bubbles in a simulation (fig. \ref{fig:sheared211}(c)), the user is
restricted to some critical number of tessellation points $M$, which
gives a limitation on accuracy even for low velocity. In fact, the
foam structure is highly non-uniform, in the sense that the bubbles
may have different sizes.
Effectively this means that every
interface has its own critical velocity and the bigger bubbles thus
introduce larger errors. This creates the following duality: to
accurately describe the process of bubble motion  it is necessary to have
the computational error as small as possible, while the error
generated near the critical velocity  takes its greatest value.
This is true for boundary or internal bubbles equally. Problems
requiring high accuracy of the solution near the external boundary
are related to the investigation of the boundary effects describing
the total phenomenological behaviour of the foam structures.

Another important remark is that the choice of the initial condition for testing
the numerical procedure here (a straight-line interface) is much more severe
than any of the instantaneous solutions reported in the Appendix. One could even
think of worse situations to test the algorithm, for example if the initial
interface is not convex or even not smooth. This may lead to super-critical
velocities and so on.

To revise and improve the existing numerical algorithm, we propose
to use another strategy for the redistribution of the tessellation
points: an equal-segment-length distribution of the tessellation
points is much more favourable \cite{Grassia1}. However, 
this strategy is not sufficient to eliminate
inaccuracy in the computation near the maximum velocity in the
steady-state regime. The reason for this is the
behaviour of the steady-state solution near the wall
(\ref{Mullin_est2b}). In fact, there exist two possible realisations
of this algorithm. One, which we have used in our computations, is to
keep the same number of tessellation points, $M$, then with
time the length between the consecutive points, $L$, will increase
significantly when $V$ is near $V_{cr}$. This leads to an effective
loss of accuracy. Another strategy would be to keep the same
distance between points during the computations. However, the
number of the points, $M$, will then increase to infinity as $V\to
V_cr$. Formally this should preserve computational accuracy but will
lead to a drastic increase in computational time and memory use, which
is unacceptable. Thus, further adaptations to the algorithm are required if high
accuracy is required, for example in the steady-state regime
with velocities near the critical value.

Taking advantage of
the auxiliary identities (\ref{10a}), (\ref{h2a}) and
(\ref{h7}), we may correct the instantaneous solution obtained
within any algorithm at any or even every time step without
time-consuming computations, as the identities are valid for any
instantaneous solution. They  also make possible further
investigation of the asymptotic behaviour of the bounded interface
solution near the ends. For example, any possible solution behaves
at the symmetry axis according to (\ref{h9}), which allows us to
tackle the error in the solution near the symmetry axis. On the
other hand, the results obtained in section \ref{sect_2.3} may allow
us to construct and implement a new numerical procedure/elements
tackling the boundary condition in a more accurate way (without
losing any near-boundary points).

Finally, as we have shown, the
identities (\ref{10a}), (\ref{h2a}) and (\ref{h7}) may be used to
probe the accuracy of computations. These indicators are extremely
helpful as they are not based on information about the exact
solution and can therefore illuminate inaccuracy of the numerical
solution without any preliminary knowledge about the exact solution
itself.

Summarizing, we have shown  that an
improvement of the numerical algorithm is highly desirable and possible.
Apart from the fact that some of the improvements have been indicated and proven in this paper, there is still an open question how to
deal with accuracy of the computations near the critical velocities and near the external boundaries.
We have also suggested possible directions for future investigation: improved
implementation of the boundary condition, creation of additional
near-boundary points. Further, to check new results related to the numerical
algorithm we need a larger set of analytical benchmarks.

\section*{Acknowledgements} The authors thank P. Grassia for his many helpful remarks on an earlier version. SJC thanks EPSRC (EP/D048397/1, EP/D071127/1) for financial support.

\appendix
\section{Appendix: A family of symmetrical instantaneous solutions}

In this section we present analytical representations of
some instantaneous symmetrical
solutions for the interface satisfying the same boundary
(\ref{Mul_11}) and symmetry (\ref{Mul_3}) conditions as Mullins'
solution.

\subsection{First example}

Let we consider the following simple combination of compatible
velocities
\begin{equation}
v_2(y)=-W^2y,\quad v_1(y)=W\sqrt{1-W^2y^2},\quad
W=\frac{V}{\sqrt{1+V^2h^2}},\label{h13}
\end{equation}
which satisfy (\ref{h7}) with $C=0$ and, as a result, can be used to
construct a symmetrical instantaneous solution. Here $W$ is the same
constant as in (\ref{h9}). Natural restrictions (\ref{h12}) for the
existence of such solution give the same estimate:
\begin{equation}
W<\frac{1}{h}, \quad \mbox{or}\quad
\frac{V}{\sqrt{1+V^2h^2}}<\frac{1}{h}, \label{cond}
\end{equation}
which holds true for any values of $V$ and $h$. The
shape of the interface is an
ellipse described by equation
(\ref{h2a}): 
\begin{equation}
x(y,t)=x(0,t)-\sqrt{1-W^2 y^2}.\label{h14}
\end{equation}

The tangential angle $\theta$ for this solution is a monotonically
decreasing function in the interval $y\in(0,h)$ and
\begin{equation}
\theta(y)\in(\theta_{min},\pi/2),\quad
\theta_{min}=\pi/2-\arcsin(Wh)>0. \label{Mul_5c}
\end{equation}

\subsection{Second example}

Let we now consider another specific instantaneous solution assuming
that $v_1=W<V$. Then the second component of the velocity satisfies
the equation
\begin{equation}
\frac{W^2}{v_2^2(y)}=-\frac{1}{2\int\limits_{0}^yv_2(\xi)d\xi}-1.\label{h8a}
\end{equation}

To find $v_2(y)$ it is more convenient to return to the
differential equation (\ref{h4}) rather than working with the nonlinear
integral equation (\ref{h8a}). After integration it takes the form
\begin{equation}
\Phi\left(\frac{v_2}{W}\right)=-Wy,\label{h16}
\end{equation}
where the odd function $\Phi$ is defined as
\begin{equation}
\Phi(\xi)=\frac{1}{2}\left(\arctan\xi+\frac{\xi}{1+\xi^2}
\right),\quad \Phi^\prime(\xi)=\frac{1}{(1+\xi^2)^2}.\label{h15a}
\end{equation}
Note that $\Phi: \mathbb{R}_+\to [0,\pi/2)$ is a monotonic function.
Moreover, one can easily obtain the constraint
$Wh<\pi/4, $
which is similar to (\ref{h12}) and (\ref{cond}).
Then the required velocity component $v_2$ can be found from:
\begin{equation}
v_2=-W\Phi^{-1}(Wy),\label{h16a}
\end{equation}
and we can finally find the complete solution using (\ref{h2}):
\begin{equation}
x(y,t)=x(0,t))+\frac{1}{W}\int\limits_0^{Wy} \Phi^{-1}(\xi)d\xi.
\label{h17}
\end{equation}

Finally, note that the tangential angle $\theta$ for this solution
is a monotonically decreasing function in the interval $y\in(0,h)$
\begin{equation}
\theta(y)\in(\theta_{min},\pi/2),\quad
\theta_{min}=\pi/2-\int\limits_{0}^{Wh}
\left(1+\Big(\Phi^{-1}(\xi)\Big)^2\right)d\xi. \label{Mul_5d}
\end{equation}
It remains only to find possible  values of the unknown constant
$W$ in order to satisfy the boundary condition (\ref{Mul_11}). The relevant
equation takes the form
\begin{equation}
\Phi^{-1}(Wh)=\sqrt{\frac{Vh}{Wh}-1}. \label{est_lats}
\end{equation}
This equation has the unique solution $W=W_*(V,h)<V$. In fact, the
left hand side is an increasing function from zero to infinity as
$Wh\to\pi/4$, whereas the right-hand side is a decreasing function
taking  values between $\infty$ when $Wh\to0$ and 0 when $Wh\to Vh$.
Additionally one can conclude from this that $Wh<\min\{Vh,\pi/4\}$,
so the restriction defined after (\ref{h15a}) always holds. In other words, this
solution, as well as that of the first example, is well-defined for
arbitrary velocity $V$ and position of the boundary $y=h$. We can
also show that $\theta_{min}$ is always positive:
\begin{equation}
\theta_{min}> \pi/2-\int\limits_{0}^{\pi/4}
\left(1+\Big(\Phi^{-1}(\xi)\Big)^2\right)d\xi=\pi/2-\int\limits_{0}^{\infty}
\left(1+\eta^2\right)\Phi^\prime(\eta)d\eta=\pi/2-\int\limits_{0}^{\infty}
\frac{d\eta}{1+\eta^2}=0.
 \label{estim}
\end{equation}
It is interesting to note that in the case $V<\!\!<1$ both of the
instantaneous solutions constructed above coincide with Mullins'
solution to within an accuracy of $O(V^2)$ for any fixed value of
$h$. On the other hand, in this case
the solution is practically (with the same accuracy) a straight
line.

\subsection{General case}

The second example above indicates how to build a wider class of symmetrical
instantaneous solutions. Let us introduce the following set
$\mathfrak{A}\subset C^2([-a,a])$, ($a>0$) of even functions,
$\psi(\xi)=\psi(-\xi)$, satisfying the following four
conditions:
\begin{equation}
\psi(\xi)=\frac{1}{2}\xi^2+O(\xi^4),\quad \xi\to0;\quad
\psi(a)\le\frac{1}{2};\quad \psi^\prime>0; \quad
\left(\frac{\psi^\prime}{\sqrt{\psi(1-2\psi)}}\right)^\prime\ge0,\quad
\xi\in(0,a).
 \label{set1}
\end{equation}
Note that $a$ may differ from function to function, but it is necessary that for every function there exists some $a>0$ for which all conditions (\ref{set1})
hold. For some functions it may happen that $a=\infty$.

For example, the following three functions belong to the set $\mathfrak{A}$:
\begin{equation}
\psi_1(\xi)=\frac{1}{2}\sin^2\xi,\quad \psi_2(\xi)=\frac{1}{2}\xi^2,\quad
\psi_3(\xi)=\int\limits_0^{\xi} \Phi^{-1}(\zeta)d\zeta=
\frac{1}{2}\left(1-\frac{1}{1+\big[\Phi^{-1}(\xi)\big]^2}\right),
 \label{examples}
\end{equation}
 with $a_1=\pi/2$, $a_2=1$ and $a_3=\infty$ respectively.
These three functions have been collected from  Mullins' solution and two previous examples.
Thus, the set $\mathfrak{A}$ is not empty.

Using any function from this set we can construct a symmetrical instantaneous
solution with velocity
components in the form:
\begin{equation}
v_2(y)=-W\psi^\prime(Wy),\quad
v_1(y)=W\psi^\prime(Wy)\sqrt{\frac{1-2\psi(Wy)}{2\psi(Wy)}},
 \label{set2}
\end{equation}
that identically satisfies equation (\ref{h7}) with $c=0$. Then the
unknown constant $W$ should be taken to be of the form $W=W_*(Vh)/h$
where $W_*(Vh)>0$ is a solution of
the implicit equation:
\begin{equation}
\frac{\psi^\prime(W_*)}{\sqrt{2\psi(W_*)(1-2\psi(W_*))}}=\frac{Vh}{W_*}.
 \label{set3}
\end{equation}

Because of the last condition in (\ref{set1}), there may exist only
one solution of this equation. If, in addition, the left-hand side
of (\ref{set1}) tends to infinity as $W_*\to a$, then the solution
always exists and $W_*<a$. However, if the left-hand side of
(\ref{set1}) tends to a finite value $L>0$ as $W_*\to a$, then the
solution exists only under the additional condition
\begin{equation}
Vh<La.
 \label{set4}
\end{equation}
One can check that for Mullins' solution
 $L=1$ and (\ref{set4}) coincides with (\ref{Mul_5a}). For the other two cases previously discussed above, we have $L=\infty$ so the solution of
the implicit equation (\ref{set3}) always exists and no solvability condition (\ref{set4}) is needed in these cases.

To reconstruct the complete symmetrical instantaneous solution based on (\ref{set2}) it is enough to substitute it in
 (\ref{10}), (\ref{10a}) and (\ref{h2a}).

Note that in the case $V<\!\!<1$,
the solution to (\ref{set3}) gives $W_*\sim V$, as one can
conclude from the first part of (\ref{set1}). This means that any instantaneous
solution differs negligibly from the Mullins' solution for
small values of the velocity $V$.

It remains to investigate two important constraints (\ref{h12}). Taking into account that
\[
\int_0^hv_2(\xi)d\xi=-\psi(Wh),\quad
\int_0^hF(\xi)d\xi=\frac{1}{2}\left(\mbox{arcsin} \big(4\psi(Wh)-1\big)+\frac{\pi}{2}\right),
\]
then the two constraints (\ref{h12}) are equivalent in this case and  correspond to $\psi(Wh)\le 1/2$, which coincides with the second part of (\ref{set1}).

In fact, the third condition, $\psi^\prime>0$, from (\ref{set1}) was
never used. We added it only to have a convex instantaneous
solution; without it, we can construct non-convex interfaces.

\subsection{Integral measures}
\label{app:notate}

Here we define explicitly the integral measures of problem quantities, used for assessing the accuracy of the computations, and their relative errors.

\noindent
After (\ref{10a}):
\begin{equation}
\mbox{intF} = -\int\limits_{0}^y F(\xi)d\xi;
\quad \mbox{relative error} = -\left(\int\limits_{0}^y F(\xi)d\xi \right)/ \left( \theta(y)-\frac{1}{2} \pi\right) -1.
\label{10a2}
\end{equation}
After (\ref{h2a}):
\begin{equation}
 \mbox{intv} =\int\limits_{0}^y\frac{v_2(\xi)}{v_1(\xi)}d\xi;
\quad  \mbox{relative error} = \left(\int\limits_{0}^y\frac{v_2(\xi)}{v_1(\xi)}d\xi \right)/ \left(x(0)-x(y)\right) -1.
\label{h2a2}
\end{equation}
After (\ref{h7}):
\begin{equation}
\mbox{intv2} =\int\limits_{0}^y v_2(\xi)d\xi;
\quad \mbox{relative error} = \left(\int\limits_{0}^y v_2(\xi)d\xi \right)/
 \left(\frac{v_2^2(y)}{2(v_2^2(y)+v_1^2(y))}\right) -1.\label{h72}
\end{equation}

\end{document}